\documentclass[a4paper,11pt]{article}

 \usepackage{amsfonts,amsthm,color,amssymb}
 \usepackage{latexsym,epsf}

\newcommand{\eq}{\begin{equation}}
\newcommand{\en}{\end{equation}}
\newcommand{\eqa}{\begin{eqnarray}}
\newcommand{\ena}{\end{eqnarray}}

\newcommand{\ha}{\hat{a}}
\newcommand{\hb}{\hat{b}}
\newcommand{\hc}{\hat{c}}
\newcommand{\hd}{\hat{d}}
\newcommand{\hi}{\hat{I}}

\newcommand{\0}{|0\rangle}
\newcommand{\1}{|1\rangle}

\newcommand{\2}{|00\rangle}
\newcommand{\3}{|01\rangle}
\newcommand{\4}{|10\rangle}
\newcommand{\5}{|11\rangle}

\begin{document}

\setlength{\unitlength}{1mm} \thispagestyle{empty}

\vspace*{0.1cm}


\begin{center}
{\small \bf  Quantum Algebras Associated With Bell States \\[2mm]}

\vspace{.5cm}

 Yong Zhang${}^a$\footnote{yzhang@nankai.edu.cn}, Naihuan Jing
${}^{bc}$\footnote{jing@math.ncsu.edu} and Mo-Lin
Ge${}^a$\footnote{geml@nankai.edu.cn}\\[.1cm]

 ${}^a$ Theoretical Physics Division, Chern Institute of Mathematics \\
  Nankai University, Tianjin 300071, P. R. China\\[0.1cm]

 ${}^b$ School of Mathematical Sciences, South China University
of Technology\\ Guangzhou, Guangdong 510641, P. R. China
  \\[0.1cm]

${}^c$
Department of Mathematics, North Carolina State University\\
Raleigh, NC 27695-8205, USA\\[0.3cm]

\end{center}

\vspace{0.2cm}

\begin{center}
\parbox{12cm}{
\centerline{\small  \bf Abstract} \small \noindent

The antisymmetric solution of the braided Yang--Baxter equation
called the Bell matrix becomes interesting in quantum information
theory because it can generate all Bell states from product states.
In this paper, we study the quantum algebra through the FRT
construction of the Bell matrix.  In its four dimensional
representations via the coproduct of its two dimensional
representations, we find algebraic structures including a
composition series and a direct sum of its two dimensional
representations to characterize this quantum algebra. We also
present the quantum algebra using the FRT construction of
Yang--Baxterization of the Bell matrix.

 }

\end{center}

\vspace*{10mm}
\begin{tabbing}

\end{tabbing}

\newpage

 \section{Introduction}

It is well known that the Bell states are maximally entangled states
and they play key roles in various topics of quantum information
\cite{nielsen} such as quantum entanglements \cite{werner}, quantum
cryptography \cite{ekert} and quantum teleportation \cite{bennett}.
The observation that all the Bell states can be generated by the
Bell matrix motivates a series of papers in the recent study: The
Bell matrix is identified with a universal quantum gate \cite{dye,
kauffman}; Yang--Baxterization of the Bell matrix \cite{yong1,yong2}
is exploited to derive the Hamiltonian determining the evolution of
the Bell states; the braid teleportation configuration
\cite{yong3,yong4} in terms of the Bell matrix is a sort of
algebraic structure underlying quantum teleportation.

The Bell matrix is a solution of the braided Yang--Baxter equation
without spectral parameters \cite{yang, baxter}, i.e., it is a
unitary braid representation which has been used to calculate the
Jones polynomial \cite{wang}. But it does not get involved in the
previous study of those topics using solutions of the Yang--Baxter
equation such as integral models \cite{faddeev} in some two
dimensional quantum field theories and statistical physics. The Bell
matrix has eight non-zero matrix entries as eight-vertex models
\cite{baxter} do. Matrix entries of symmetric solutions of
eight-vertex models \cite{baxter} are either positive or zero and
they are explained as the Boltzman weights in statistical physics,
but the Bell matrix has negative matrix entries which can not be the
Boltzman weights. Here for convenience we call the Bell matrix
antisymmetric solution of eight-vertex models since it also has
eight non-vanishing matrix entries and it is non-triangular and
non-singular\footnote{In the old time the term eight-vertex model
was a pronoun meaning the original one solved by Baxter
\cite{baxter}.}.

In the literature there is a long history of dealing with
eight-vertex models \cite{baxter, faddeev}, related the Sklyanin
algebra \cite{sklyanin1,sklyanin2} and some relevant extensions
including Potts models \cite{faddeev, bajhanov}. Solutions of the
Yang--Baxter equation with the spectral parameter, i.e., symmetric
solutions of eight-vertex models, have double-period elliptic
functions \cite{baxter, faddeev} as matrix entries. However, they
are nothing with the recent study in quantum information. As a
matter of fact, we find standard methodologies for six-vertex models
\cite{faddeev,kassel} helpful for the purpose of exploring
applications of the Bell matrix to quantum information. For example,
Yang--Baxterization of the Bell matrix \cite{yong1,yong2} has matrix
entries in terms of single-period trigonometric functions which
characterize six-vertex models.

In view of the fact that a new quantum group can be found via a
`non-standard' braid group representation \cite{jing}, in this paper
we study the quantum algebra via the $FRT$ recipe \cite{rtf} which
was devised for six-vertex models \cite{faddeev,kassel}, i.e., the
generators of this algebra satisfying the $\check{R}TT$ relation
where the symbol $R$ denotes the Bell matrix and symbol $T$ denotes
a matrix collecting all generators. We find that there exists a
composition series in its four dimensional representation, and shows
that a product of its two dimensional representations can be
decomposed into a direct sum of its two dimensional representations.
Also, we have another quantum algebra through the $FRT$ recipe on
Yang--Baxterization of the Bell matrix.

The plan of this paper is organized as follows. Section 2 introduces
the Bell states, Bell matrix and Yang--Baxterization of the Bell
matrix. Section 3 derives the quantum algebra using the
$\check{R}TT$ relation of the Bell matrix. Section 4 lists its two
dimensional representations and presents two examples for its four
dimensional representations to show characteristic properties of
this quantum algebra. Section 5 sketches the quantum algebra using
the $\check{R}TT$ relation of Yang--Baxterization of the Bell
matrix. The last section has open problems to be solved.

\section{Bell states, Bell matrix and Yang--Baxterization}

The two-dimensional unit matrix is denoted by $1\!\! 1_2$, the Pauli
matrices $\sigma_x, \sigma_y, \sigma_z$ have the conventional forms,
and the symbols $\sigma_+, \sigma_-$ are given by \eq \sigma_+=\frac
1 2 (\sigma_x+i \sigma_y)= \left(\begin{array}{cc}
   0  &  1 \\
   0  &  0
  \end{array} \right),\,\, \sigma_- =\frac 1 2
  (\sigma_x-i \sigma_y)= \left(\begin{array}{cc}
   0  &  0 \\
   1 &  0
  \end{array} \right),
 \en which are nilpotent operators satisfying $\sigma_+^2=\sigma_-^2=0$.
Denote the bases of a two dimensional vector space over the complex
field ${\mathbb C}$ in terms of the Dirac notation $|0\rangle,
|1\rangle$ which form product bases $|ij\rangle, i,j=0,1$ of a four
dimensional vector space over $\mathbb C$. Four orthonormal Bell
states have the forms, \eq \label{be} |\psi_\pm\rangle= \frac 1
{\sqrt{2}} (|0 0\rangle \pm |1 1\rangle), \qquad
|\phi_\pm\rangle=\frac 1 {\sqrt{2}} (| 1 0 \rangle \pm |0 1
\rangle). \en which are transformed to each other under local
unitary transformations, \eq
 |\psi_-\rangle=(1\!\! 1_2\otimes \sigma_z) |\psi_+\rangle,\,\,
  |\phi_+\rangle=(1\!\! 1_2\otimes \sigma_x) |\psi_+\rangle,\,\,
  |\phi_-\rangle=(1\!\! 1_2\otimes -i\sigma_y) |\psi_+\rangle.
\en

In the literature \cite{yong1, yong2,yong3,yong4}, there are two
types of the Bell matrix: $B_+$ and $B_-$ given by
 \eq \label{bell}
 B_+=\frac 1 {\sqrt 2}\left(
 \begin{array}{cccc}
 1 & 0 & 0 & 1 \\
 0 & 1 & 1 & 0 \\
 0 & -1 & 1 & 0 \\
 -1 & 0 & 0 & 1 \\
 \end{array} \right), \qquad
 B_-=\frac 1 {\sqrt 2}\left(
 \begin{array}{cccc}
 1 & 0 & 0 & 1 \\
 0 & 1 & -1 & 0 \\
 0 & 1 & 1 & 0 \\
 -1 & 0 & 0 & 1 \\
 \end{array} \right)
\en which have formalisms of exponential functions,  \eq
 B_+=e^{i\frac {\pi} 4 (\sigma_y\otimes \sigma_x)},\qquad
 B_-=e^{i\frac {\pi} 4 (\sigma_x\otimes \sigma_y)}
\en with interesting properties:
 \eq B_\pm^4=-1\!\! 1_4, \qquad B_\pm^8=1\!\! 1_4, \qquad
 B_\pm=\frac 1 {\sqrt 2} (1\!\! 1_4 + B_\pm^2). \en
In terms of the Bell matrices and product bases, the Bell states can
be generated in the way  \eqa
  & B_+|00\rangle=|\psi_-\rangle,\,\, B_+|1 1\rangle=|\psi_+\rangle,
 \qquad B_+| 0 1 \rangle=|\phi_-\rangle,\,\, B_+|1 0\rangle
 =|\phi_+\rangle, \nonumber\\
  & B_-|00\rangle=|\phi_-\rangle,\,\, B_-|11\rangle=|\phi_+\rangle,
 \qquad B_-|01\rangle=|\psi_+\rangle, B_-|10\rangle=-|\psi_-\rangle. \ena

 The Bell matrix $B_\pm$ forms a unitary braid representation, i.e.,
 satisfying the Yang--Baxter equation without the spectral
 parameter,  see \cite{yong3,yong4} for the proof. Via Yang--Baxterization
 \cite{yong1, yong2},  the corresponding $B_\pm(x)$ matrix which is
 a solution of the Yang--Baxter equation with the spectral parameter
 $x$ has the form
  \eq \label{yb} B_\pm(x) =\frac 1 {\sqrt {2(1+x^2)}}
 \left(\begin{array}{cccc}
 1+x & 0 & 0 & q(1-x) \\
 0 & 1+x & \pm(1-x) & 0 \\
 0 & \mp(1-x) & 1+x & 0 \\
 -q^{-1} (1-x) & 0 & 0 & 1+x \end{array}\right),  \en where
$B_\pm=B_\pm(0)|_{q=1}$, $q$ called the deformation
parameter\footnote{Note that the $B_\pm(x)$ matrix satisfies the
free fermion condition \cite{fanwu} and therefore is a special case
of the free fermion $\check{R}$-matrix known from
\cite{felderhof}.}.

With new variables of angles $\theta$ and $\varphi$, \eq
\label{belltypetran} \cos\theta=\frac 1 {\sqrt{1+x^2}}, \qquad
 \sin\theta=\frac x {\sqrt{1+x^2}}, \qquad
 q=e^{-i\varphi},
 \en
we rewrite the $B_\pm(x)$ matrix into an exponential formalism, \eq
B_\pm(\theta)=\cos(\frac \pi 4-\theta)+ 2\,i\,
 \sin(\frac \pi 4-\theta)\,H_{\pm}=e^{i (\frac \pi
 2-2\,\theta) H_\pm} \en
where the symbol $H_\pm$ called the Hamiltonian \cite{yong1,yong2}
is given by \eqa
 & & H_+=\frac 1 2 \sigma_{n_1}\otimes
\sigma_{n_2}, \qquad H_-=\frac 1
2\sigma_{n_2}\otimes \sigma_{n_1}, \nonumber\\
&& \sigma_{n_1} = \sigma_+ e^{-\frac i 2 (\varphi+\pi)}+\sigma_-
e^{\frac i 2 (\varphi+\pi)},\qquad \sigma_{n_2} = \sigma_+ e^{-\frac
i 2 \varphi}+\sigma_- e^{\frac i 2 \varphi}. \ena The time evolution
of the Bell states under the Hamiltonian $H_\pm$ with the time
variable $\theta$ is determined by the unitary
$B_\pm(\theta)$-matrix, i.e., \eqa & & B_\pm(\theta) |0 0\rangle
=\cos(\frac \pi 4-\theta)
 |00\rangle- e^{i\varphi}\sin(\frac \pi 4-\theta) |1
 1\rangle,\nonumber\\
 & & B_\pm(\theta) |1 1\rangle =e^{-i\varphi} \sin(\frac \pi 4-\theta)
 |0 0\rangle +\cos(\frac \pi 4-\theta) |1 1\rangle, \nonumber\\
 & & B_\pm(\theta) |0 1\rangle =\cos(\frac \pi 4-\theta) |0 1\rangle \mp
\sin(\frac \pi 4-\theta) |1 0\rangle,\nonumber\\
 & & B_\pm(\theta) |1 0\rangle = \pm \sin(\frac \pi 4-\theta) |0
 1\rangle) +\cos(\frac \pi 4-\theta)|1 0\rangle.
  \ena

 \section{Quantum algebra associated with the Bell matrix}

Let us start with the $\check{R}_\omega$-matrix, a solution of the
Yang--Baxter equation without spectral parameter \cite{yong1,
yong2}, \eq \label{symant}
\check{R}_\omega=\left(\begin{array}{cccc}
1 & 0 & 0 & q \\
0 & 1 & 1 & 0 \\
0 & \omega & 1 & 0 \\
\omega\, q^{-1} & 0 & 0  & 1
\end{array}\right), \qquad \omega=\pm 1,\qquad q\neq 0,\,\,\,q\in {\mathbb
C} \en where the $\check{R}_{-1}$-matrix is a deformation of the
Bell matrix $B_+$, and the $\check{R}_1$-matrix is a deformation of
a symmetric solution of eight-vertex models.

In this section, we set up a quantum algebra ${\cal A}_{-1}$ via the
$\check{R}TT$ relation where the $R$-matrix is $\check{R}_{-1}$. The
$T$-matrix has  non-commutative operators $\hat{a}$, $\hb$, $\hc$
$\hd$ as its matrix entries, and its tensor product $T\otimes T$ has
the form \eq T=\left(\begin{array}{cc}
   \hat{a}  &  \hb  \\
   \hc  &  \hd   \\
 \end{array}\right),\qquad  T\otimes T=\left(\begin{array}{cccc}
\hat{a} \hat{a} & \hat{a} \hb & \hb \hat{a} & \hb \hb \\
\hat{a} \hc & \hat{a} \hd & \hb \hc  & \hb \hd \\
\hc \hat{a} & \hc \hb & \hd \hat{a} & \hd \hb \\
\hc \hc & \hc \hd & \hd \hc  & \hd \hd
\end{array}\right)
\en where the tenor symbol $\otimes$ in every matrix entry of
$T\otimes T$ has been omitted for convenience. The $\check{R}TT$
relation: $\check{R}(T\otimes T)= (T\otimes T) \check{R}$ leads to a
quantum algebra with the generators $\hat{a}, \hb, \hc, \hd$
satisfying algebraic relations, \eqa \label{algebra}
 \ha \ha &=&  \hd \hd, \,\, \ha \hb = q \hd \hc,\,\,\hb \hb =  \omega q^2 \hc
 \hc,\,\,\ha \hc =q^{-1} \hd \hb, \nonumber\\
 \ha \hd &=& \hd \ha,\,\, \hb \ha = \omega q \hc \hd, \,\,
 \hb \hc = \omega \hc \hb, \,\,\hc \ha = \omega q^{-1} \hb \hd. \ena

In this paper, however, we will study the quantum algebra $\cal
A_\omega$ generated by four generators $\ha,\hb,\hc,\hd$ satisfying
the algebraic relations (\ref{algebra}) except the following two
equations, \eq
 [F_1] \equiv \hc\ha-\omega q^{-1} \hb\hd =0, \qquad [F_2]\equiv \ha\hc -
 q^{-1} \hd \hb =0
\en because we can derive the following equations, \eqa & &
 \ha  [F_2] =[F_1]\ha=0, \qquad \hd [F_2] = [F_1] \hd =0,
 \nonumber\\
 & & \hb [F_1] = [F_2] \hb =0, \qquad \hc[F_1] =[F_2] \hc=0
\ena in terms of the remaining six algebraic relations. As any one
of four generators $\ha,\hb,\hc,\hd$ is invertible or nipotent,
$[F_1]$ and $[F_2]$ will be obviously vanishing, that is to say: in
these cases, the ${\cal A}_\omega$ algebra is equivalent to the
original quantum algebra derived from the $\check{R}TT$ relation.

With the help of a rescaling $q \hc \to \hc$, i.e., the deformation
parameter $q$ to be absorbed into a new generator $\hc$, the quantum
algebra $\cal A_\omega$ is generated by $\ha,\hb,\hc,\hd$ satisfying
algebraic relations, \eq
 \label{commu}
 \hat{a} \ha =  \hd \hd, \,\, \ha \hd = \hd \ha,\,\,
  \hb \hb = \omega  \hc \hc, \,\, \hb \hc = \omega \hc \hb,
  \,\,\ha \hb = \hd \hc,\,\, \hb \ha =\omega\hc \hd.
 \en
The quantum algebra ${\cal A}_1$, i.e.,  $\omega=1$, presents a
known quantum algebra obtained from the $\check{R}TT$ relation of
symmetric solutions of eight-vertex models. But the quantum algebra
${\cal A}_{-1}$, i.e., $\omega=-1$ is very attractive because the
Bell matrix becomes interesting only in the recent study of quantum
information.

The quantum algebra ${\cal A}_{-1}$ has interesting quotient
algebras. As the generator $\hat{a}$ is a complex scalar denoted by
$\ha=p \hi$ with the unit $\hi$,  the algebra ${\cal A}_{-1}$ is
reduced
 to an algebra generated by $\ha, \hc, \hd$ satisfying
 algebraic relations,
 \eq
 \ha=p\hi, \qquad \hd^2 =p^2 \hi, \qquad \hc \hd =-\hd
 \hc, \qquad p\neq 0, \,\, p\in {\mathbb C},
 \en
where we define a composition operator $\hb$ to denote the operator
product $\hc\hd$, i.e., $\hb = p^{-1} \hd \hc$, proved to satisfy
$\hb^2 =-\hc^2$ and $\hb\hc =-\hc\hb$. Another interesting quotient
algebra is to require $\hb, \hc$ to describe fermions, i.e., they
satisfying $\hb^2=\hc^2=0, \hb\hc=-\hb\hc$. Specially, as $\hb=\hc$,
they are the same fermion.

The quantum algebra ${\cal A}_{-1}$ has one dimensional
representations over the complex field $\mathbb C$: $\hb=\hc=0$ and
$\ha, \hd$ are complex numbers satisfying $\ha^2=\hd^2$, while it
also has one dimensional representation over the field including
non-commutative Grassman numbers: $\hb, \hc$ are Grassman numbers
and $\ha, \hd$ are complex numbers.

\section{Two and four dimensional representations of ${\cal A}_{-1}$}

In this section, we list two dimensional representations over
$\mathbb C$ for the quantum algebra ${\cal A}_{-1}$ and explore
interesting algebraic structures underlying its four dimensional
representations over $\mathbb C$ via the coproduct \cite{kassel} of
the generators of ${\cal A}_{-1}$.

\subsection{Two dimensional representations}

The quantum algebra ${\cal A}_{-1}$ has two subalgebras formed by
$\ha, \hd$ and $\hb, \hc$ respectively. Hence it is convenient to
require either $ \ha$ ($\hd$) or $\hb$ ($\hc$) to be a diagonal
matrix in a two-dimension representation of ${\cal A}_{-1}$. Note
that the following two-dimension representations of ${\cal A}_{-1}$
will be found to satisfy $[F_1]=[F_2]=0$.

As the generator $\ha$ has a non-vanishing eigenvalue $\lambda$ with
two degenerate eigenvectors, the generators $\ha, \hb, \hc, \hd$
have the following two dimensional representations over the complex
field $\mathbb C$,
 \eqa \ha &=& \left(\begin{array}{cc}
   \lambda & 0 \\
   0  &  \lambda
  \end{array} \right),\qquad \hd = \left(\begin{array}{cc}
   \alpha  &  \beta \\
   \gamma  &  -\alpha
  \end{array} \right), \qquad  \alpha^2 + \beta \gamma
  =\lambda^2, \lambda\neq 0, \lambda\in {\mathbb C}  \nonumber\\
   \hc &=& \left(\begin{array}{cc}
   c_1  &  c_2 \\
   c_3  &  -c_1
  \end{array} \right), \qquad
  2 c_1 \alpha = -c_3 \beta -c_2 \gamma,\,\,
  \alpha,\beta,\gamma,c_1,c_2,c_3\in {\mathbb C}
 \ena
where the generator $\hb$ is determined by $\hb=\lambda^{-1}\hd\hc$.
In this representation, taking $\alpha=0, \beta=\gamma=\lambda$ and
$c_1=\mu, c_2=c_3=0$ leads to a representation in terms of the unit
matrix $1\!\! 1_2$ and Pauli matrices,
 \eq \label{scalar} \ha=\lambda 1\!\! 1_2,\,\, \hd = \lambda \sigma_x,\qquad
 \hb=-i \mu  \sigma_y, \,\, \hc=\mu \sigma_z,\qquad
  \lambda, \mu\in{\mathbb C} \en
while taking $\alpha=\lambda=1, \beta=\gamma=0$  and $c_1=0,
c_2=c_3=1$ gives another interesting representation,
 \eq \ha=1\!\! 1_2,\,\, \hd=\sigma_z,\qquad \hb=i \sigma_y,\,\,\hc=\sigma_x.
 \en

As the generator $\ha$ has two distinct complex eigenvalues,
$\lambda_1\neq \lambda_2$, the two dimensional representation for
the generators $\ha, \hb, \hc, \hd$ over the complex field $\mathbb
C$ is obtained to be
   \eq \ha =
\left(\begin{array}{cc}
   \lambda_1 & 0 \\
   0  &  \lambda_2
  \end{array} \right),\hd = \epsilon\left(\begin{array}{cc}
   \lambda_1 & 0 \\
   0  &  - \lambda_2
  \end{array} \right),  \hb=\epsilon \left(\begin{array}{cc}
   0 & c_2 \\
   -c_3  &  0
  \end{array} \right), \hc=\left(\begin{array}{cc}
   0 & c_2 \\
   c_3  &  0
  \end{array} \right), \en
where the parameter $\epsilon$ satisfies $\epsilon^2=1$. As $\hb,
\hc$ represent the same fermion, i.e., $\epsilon=1, c_2=1, c_3=0$,
we have the two dimensional representation, \eq \label{fermion}
 \hb=\hc=\left(\begin{array}{cc}
   0  &  1 \\
   0  &  0
  \end{array} \right),  \qquad  \ha =
\left(\begin{array}{cc}
   \lambda_1 & 0 \\
   0  &  \lambda_2
  \end{array} \right), \qquad  \hd = \left(\begin{array}{cc}
   \lambda_1 & 0 \\
   0  &  - \lambda_2
  \end{array} \right).
\en

Now we study the two dimensional representation in which the
generator $\hb$ is a diagonal matrix. If $\hb$ has an eigenvalue
with two degenerate eigenvectors, i.e., it is a scalar operator, the
generator $\hc$ has to be vanishing.  As $\hb$ has two distinct
eigenvalues $p_1, p_2$ with two eigenvectors $\vec{v}_1, \vec{v}_2$,
$p_1=-p_2$ has to be satisfied because $\hc \vec{v}_1$ is an
eigenvector of $\hb$ with the eigenvalue $-p_1$. Hence  the
generators $\ha,\hb,\hc,\hd$ have a non-vanishing two dimensional
representation over $\mathbb C$, \eqa
 \hb &=& p \sigma_z, \qquad \hc=\left(\begin{array}{cc}
0   & - p^2    \\
1   &  0
\end{array}  \right),\qquad p\neq 0, \nonumber\\
 \ha &=& \left(\begin{array}{cc}
  \alpha & p^2 \beta \\
  \beta  &  \alpha
 \end{array} \right)=\alpha 1\!\! 1_2+ \beta \left(\begin{array}{cc}
  0 & p^2 \\
  1  &  0
 \end{array} \right),\nonumber\\
 \hd &=& \left(\begin{array}{cc}
p \beta & p \alpha \\
 p^{-1} \alpha  &  p \beta
 \end{array} \right)= p \beta 1\!\! 1_2 + p^{-1} \alpha \left(\begin{array}{cc}
  0 & p^2 \\
  1  &  0
 \end{array} \right),\qquad \beta\neq 0
\ena where $\beta=0$ gives an example for the case that $\ha$ is a
scalar.

\subsection{Four dimensional representation: composition series}

The coproduct \cite{kassel} is a linear map $\Delta$ from the vector
space $V$ to its tensor product $V\otimes V$ satisfying the
coassociativity axiom $(\Delta\otimes Id)\circ \Delta=(Id\otimes
\Delta)\circ \Delta$ where $Id$ is an identity map. The coproducts
of the generators $\ha, \hb, \hc, \hd$ have the forms \eqa
 \Delta(\ha) &=& \ha \otimes \ha^\prime  + \hb \otimes \hc^\prime , \qquad
  \Delta(\hb)=\ha\otimes \hb^\prime + \hb\otimes \hd^\prime , \nonumber\\
 \Delta(\hc) &=& \hc \otimes \ha^\prime  + \hd \otimes \hc^\prime , \qquad
 \Delta (\hd)=\hc \otimes \hb^\prime  + \hd \otimes \hd^\prime
 \ena
where the generators $\ha^\prime, \hb^\prime, \hc^\prime,
\hd^\prime$ satisfy the quantum algebra ${\cal A}_{-1}$ and play the
same roles as $\ha, \hb, \hc, \hd$, respectively. It is easy to
prove the above coproducts to satisfy algebraic relations of the
quantum algebra ${\cal A}_{-1}$, for example, \eq
 \label{coproduct}
\Delta(\hat{a}) \Delta(\ha) =  \Delta(\hd)\Delta(\hd), \qquad
\Delta(\ha) \Delta(\hd) = \Delta(\hd)\Delta(\ha)
 \en
 and also $\Delta([F_1])=\Delta([F_1])=0$ as $[F_1]=[F_2]=0$.

Here we derive four dimensional representations $\underline{4}$ of
 ${\mathcal A}_{-1}$ via its coproduct structure in terms of its known
 two dimensional representations and denote this sort of four dimensional
 representation by $\underline{4}=\underline{2}\otimes \underline{2}$.
 The representation $\underline{4}$ is defined by the coproduct map
 from the generators $\ha,\hb,\hc,\hd$ to their coproducts
$\Delta(\ha),\Delta(\hb)$, $\Delta(\hc),\Delta(\hd)$ in the way
 \eq \ha|ij\rangle \equiv \Delta(\ha) |ij\rangle,\,\,
 \hb|ij\rangle \equiv \Delta(\hb) |ij\rangle,\,\,
 \hc|ij\rangle \equiv \Delta(\hc) |ij\rangle,\,\,
 \hd|ij\rangle \equiv \Delta(\hd) |ij\rangle \en
where $\0, \1$ are the bases of $\underline{2}$ and $|ij\rangle,
i,j=0,1$ are the bases of $\underline{4}$. And these four
dimensional representations of ${\cal A}_{-1}$ satisfy
$[F_1]=[F_2]=0$, too.

In the following, we present two examples for its four dimensional
representations. In the first one, we exploit the two dimensional
representation (\ref{fermion}) with $\lambda_1$ relabeled to be $1$
and $\lambda_2$ to be $\lambda$, i.e., \eqa & &\ha\0=\0, \,\,
\ha\1=\lambda\1, \qquad
\hd\0=\0,\,\,\hd\1=-\lambda\1,\nonumber\\
 & & \hb\0=0, \,\,\,\,\,\, \hb\1=\0,\qquad  \hc \0=0,\,\, \hc\1=\0 \ena
and the two dimensional representation for the generators
$\ha^\prime, \hb^\prime, \hc^\prime, \hd^\prime$,
 \eqa & &\ha^\prime\0=\0, \,\, \ha^\prime\1=\lambda^\prime\1,
\qquad
\hd^\prime\0=\0,\,\,\hd^\prime\1=-\lambda^\prime\1,\nonumber\\
 & & \hb^\prime\0=0, \,\,\,\,\,\, \hb^\prime\1=\0,
 \qquad  \hc^\prime \0=0,\,\, \hc^\prime\1=\0 \ena

This four dimensional representation $\underline{4}$ has a
composition series over $\mathbb{C}$,
 \eqa \label{cs}  \{0\} \subset  \{\2\} \subset \{\2, \3\}
\subset \{\2, \3\,\4\} \subset \{\2, \3, \4\, \5\},  \ena where the
set $\{0\}$ consists of zero element and the set $ \{\2, \3\}$ can
be replaced by $\{\2, \4\}$. An ascending chain of
subrepresentations is called a composition series if the successive
quotients are irreducible representations. We will see that each
term in our composition series is an indecomposable representation
(i.e. not a direct summand) and all the nontrivial term is
reducible. Namely, the four dimensional representation has
irreducible subrepresentations but it is not completely reducible.

Let us explain this in detail. The vector space $\{\2\}$ forms an
one-dimension irreducible subrepresentation of ${\cal A}_{-1}$, \eq
 \ha\2 = \2,  \hb\2 = 0, \hc\2 = 0,  \hd\2 = \2;
\en the vector space $\{\2, \3\}$ forms a two-dimension
subrepresentation of ${\cal A}_{-1}$, \eq \ha\3 =
\lambda^\prime\3,\hb\3 = \2, \hc\3 = \2, \hd\3 = -\lambda^\prime\3;
\en and the vector space $\{\2, \4\}$ forms another two dimensional
subrepresentation of ${\cal A}_{-1}$, \eq \ha\4 =\lambda\4,\hb\4
=\2, \hc\4 = \2, \hd\4 = -\lambda\4; \en and so the vector space
$\{\2, \3\, \4\}$ forms a three-dimension subrepresentation for the
algebra ${\cal A}_{-1}$.

See Figure 1, every horizontal line represents a state in the four
dimensional representation $\underline{4}$ and every line with an
oriented arrow denotes a transition between different states which
is caused by the action of a generator.

In the four dimensional representation $\underline{4}$=$\{\2, \3\,
\4, \5\}$, the actions of all generators on $|11\rangle$ are linear
combinations of $|ij\rangle, i,j=0,1$,
 \eqa & & \ha\5 =\2+ \lambda\lambda^\prime\5,\qquad
  \hb\5 =-(\lambda^\prime\3-\lambda\4),
  \nonumber\\ & & \hc\5 = \lambda^\prime\3-\lambda \4,\qquad
  \hd\5 = \2+\lambda\lambda^\prime \5,
\ena and so it is impossible for $\underline{4}$ to have completely
reducible representations. Similarly one sees from Figure 1 that
$\{\2\}$ is the only nontrivial subrepresentation of $\{\2, \3\}$,
and hence the representations $\{\2, \3\}$, $\{\2, \4\}$, $\{\2,
\3\, \4\}$ and $\{\2, \3, \4\, \5\}$ are reducible and
indecomposable representations.

As $\lambda\lambda^\prime\neq 1$, we obtain the second common
eigenvector $|\psi\rangle$ of the generators $\ha, \hd$ which have
the first common eigenvector $|00\rangle$, \eq |\psi\rangle
=-|00\rangle + (1-\lambda\lambda^\prime) |11\rangle, \qquad  \ha
|\psi\rangle =\hd |\psi\rangle = \lambda\lambda^\prime |\psi\rangle.
\en In the vector space spanned by $|01\rangle, |10\rangle$, a
series of vectors $|\phi_n\rangle$ given by \eq
|\phi_n\rangle=(\lambda^\prime)^n |01\rangle -\lambda^n
|10\rangle,\qquad  \ha |\phi_n\rangle=|\phi_{n+1}\rangle,\,\,
 \hd |\phi_n\rangle=-|\phi_{n+1}\rangle,\,\,   n\in {\mathbb N}
 \en
form a four-dimension representation of the quantum algebra
 ${\cal A}_{-1}$ together with the vectors $|\psi\rangle$ and
 $|00\rangle$,
 \eq
 \hb |\psi\rangle =-\hc |\psi\rangle=-(1-\lambda\lambda^\prime)
 |\phi_1\rangle,\qquad \hb|\phi_n\rangle = \hc|\phi_n\rangle=
  (( \lambda^\prime)^{n} - \lambda^n   ) |00\rangle,
\en where  $|00\rangle$ with any one of $|\phi_n\rangle$ forms a
two-dimension irreducible representation.

 As $\lambda\lambda^\prime=1$, we define
$|\psi_n\rangle=n |00\rangle + |11\rangle, n\in {\mathbb N}$,
satisfying \eq  \ha |\psi_n\rangle=\hd
|\psi_n\rangle=|\psi_{n+1}\rangle,\qquad
  \hb |\psi_n\rangle= - |\phi_1\rangle, \,\,
  \hc |\psi_n\rangle =|\phi_1\rangle
\en which form a four-dimension representation of ${\cal A}_{-1}$
with $|\phi_n\rangle$ and $|00\rangle$. See Figure 1 where
composition series can be easily recognized at the diagrammatical
level.

\begin{figure}
\begin{center}
\epsfxsize=12.cm \epsffile{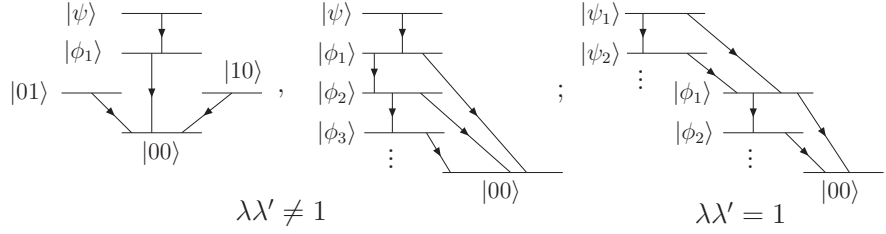} \caption{Four
dimensional representation of  ${\cal A}_{-1}$ with a composition
series (\ref{cs}).} \label{fig1}
\end{center}
\end{figure}

\subsection{Four dimensional representation:
 $\underline{2}\otimes \underline{2}=\underline{2}\oplus \underline{2}$}

In the second example, we obtain the four dimensional representation
$\underline{4}$ of ${\cal A}_{-1}$ via the coproduct construction in
terms of the two dimensional representation (\ref{scalar}) of the
generators $\ha, \hb, \hc, \hd$, \eqa & & \ha\0=\lambda\0,\,\,
  \ha\1=\lambda\1,\qquad \hd\0=\lambda\1,\,\, \hd\1=\lambda\0, \nonumber\\
 & & \hb\0=\mu\1,\,\hb\1=-\mu\0,\qquad \hc \0=\mu\0,\,\, \hc\1=-\mu\1 \ena
and the two dimensional representation of the generators
$\ha^\prime, \hb^\prime, \hc^\prime, \hd^\prime$ which is the same
as (\ref{scalar}) except that $\lambda, \mu$ are replaced by
$\lambda^\prime, \mu^\prime$. In addition, new symbols
  $z, \bar{z}, w, \bar{w}$ are introduced,
\eq z=x-i y, {\bar z}=x+i y, \qquad w= v- i u , {\bar w} =v+ i u \en
where symbols $x, y, v, u$ denote the following products, \eq
x=\lambda\lambda^\prime, y=\mu\mu^\prime,\qquad u=\lambda\mu^\prime,
v=\mu\lambda^\prime. \en

In this four dimensional representation $\underline{4}$, the
generator $\hd$ has four eigenvectors denoted by four Dirac kets
$|\chi_1\rangle, |\chi_2\rangle, |\tau_1\rangle, |\tau_2\rangle$ in
terms of Bell states $|\psi_\pm\rangle, |\phi_\pm\rangle$ (\ref{be})
given by
 \eqa & & |\chi_1\rangle= \frac 1 {\sqrt 2} ( |\psi_+\rangle+ i
 |\phi_+\rangle),\qquad
 |\chi_2\rangle= \frac 1 {\sqrt 2} ( |\psi_-\rangle - i |\phi_-\rangle),\nonumber\\
 & & |\tau_1\rangle=\frac 1 {\sqrt 2} (|\psi_-\rangle + i |\phi_-\rangle),\qquad
   |\tau_2\rangle=\frac 1 {\sqrt 2} (|\psi_+\rangle - i |\phi_+\rangle),
\ena which are related to four distinct eigenvalues of the generator
$\hd$, \eq \hd |\chi_1\rangle=z|\chi_1\rangle, \hd |\chi_2\rangle=-z
|\chi_2\rangle,\qquad \hd |\tau_1\rangle =-{\bar z}|\tau_1\rangle,
\hd |\tau_2\rangle ={\bar z}|\tau_2\rangle. \en The generator $\ha$
has an eigenvalue $z$ with two degenerate eigenvectors
$|\chi_1\rangle, |\chi_2\rangle$ and another eigenvalue $\bar z$
with two degenerate eigenvectors $|\tau_1\rangle, |\tau_2\rangle$,
i.e., \eq \ha |\chi_1\rangle
 =z|\chi_1\rangle, \ha |\chi_2\rangle = z |\chi_2\rangle,
 \qquad \ha |\tau_1\rangle={\bar z} |\tau_1\rangle,
 \ha |\tau_2\rangle ={\bar z} |\tau_2\rangle.
  \en

The Dirac kets $|\chi_1\rangle, |\tau_1\rangle$ form a two-dimension
{\em cyclic} representation  $\underline{2}$ for the generators
$\hb, \hc$: \eq \hb |\chi_1\rangle=-{\bar w} |\tau_1\rangle, \hb
|\tau_1\rangle= w |\chi_1\rangle,   \qquad \hc |\chi_1\rangle={\bar
w} |\tau_1\rangle, \hc |\tau_1\rangle= w |\chi_1\rangle, \en and the
Dirac kets $|\chi_2\rangle, |\tau_2\rangle$ form another two
dimensional {\em cyclic} representation $\underline{2}$ given by \eq
\hb |\chi_2\rangle= {\bar w} |\tau_2\rangle,
 \hb|\tau_2\rangle=-w |\chi_2\rangle,\qquad \hc|\chi_2\rangle=
{\bar w} |\tau_2\rangle,
 \hc |\tau_2\rangle=w |\chi_2\rangle. \en
 Hence we reduce the four dimensional representation
 $\underline{2}\otimes \underline{2}$ into the
 direct sum of two dimensional representations:
 $\underline{2}\otimes \underline{2}=\underline{2}\oplus
 \underline{2}$. See Figure 2 where states denoted by thick lines
 have the same eigenvalue and every irreducible representation $\underline{2}$
 is {\em cyclic}.

\begin{figure}
\begin{center}
\epsfxsize=10.cm \epsffile{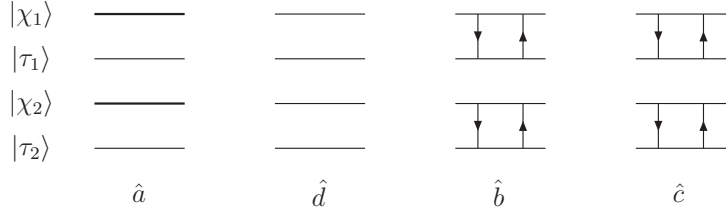} \caption{A four dimensional
representation of  ${\cal A}_{-1}$: $\underline{2}\otimes
\underline{2}=\underline{2}\oplus\underline{2}$.} \label{fig2}
\end{center}
\end{figure}

 With the bases of
 $|\chi_1\rangle, |\chi_2\rangle, |\tau_2\rangle, |\tau_1\rangle$,
 the four dimensional representations of the generators $\ha, \hb, \hc, \hd$
 are given by the following matrices,  \eqa
 & & \ha =\left(\begin{array}{cc}
  z & 0\\
  0 & {\bar z}
  \end{array}\right) \otimes 1\!\! 1_2, \qquad
 \hb =\left(\begin{array}{cc}
  0 & w\\
  {\bar w} & 0
  \end{array}\right) \otimes i\sigma_y, \nonumber\\
& & \hc = \left(\begin{array}{cc}
  0 & w\\
  {\bar w} & 0
  \end{array}\right) \otimes \sigma_x, \qquad
  \hd =\left(\begin{array}{cc}
  z & 0\\
  0 & {\bar z}
  \end{array}\right) \otimes \sigma_z.
 \ena

These four Dirac kets  $|\chi_1\rangle, |\chi_2\rangle,
|\tau_1\rangle, |\tau_2\rangle$ are found to be local unitary
transformations of the Bell states (\ref{be}):
 \eqa
 |\chi_1\rangle =(1\!\! 1_2 \otimes U_1) |\psi_+\rangle, \qquad
 |\chi_2\rangle =(1\!\! 1_2 \otimes U_2) |\psi_+\rangle, \nonumber\\
 |\tau_1\rangle =(1\!\! 1_2 \otimes U_3) |\psi_+\rangle, \qquad
 |\tau_2\rangle =(1\!\! 1_2 \otimes U_4) |\psi_+\rangle,
 \ena
where $U_1, U_2, U_3, U_4$  are unitary matrices given by \eqa
 & & U_1=\frac 1 {\sqrt 2} (1\!\! 1_2 + i \sigma_x), \qquad
 U_2 =\frac 1 {\sqrt 2} (\sigma_z -\sigma_y), \nonumber\\
 & & U_3 = \frac 1 {\sqrt 2} ( \sigma_z +\sigma_y ), \qquad
  U_4 =  \frac 1 {\sqrt 2} ( 1\!\! 1_2 - i \sigma_x )
\ena and they form a set of complete orthonormal bases for $2\times
2$ matrices, \eq tr (U_i^\dag U_j) =2 \delta_{ij},\qquad
i,j=1,2,3,4. \en Hence,  $|\chi_1\rangle, |\chi_2\rangle,
|\tau_2\rangle, |\tau_1\rangle$ are maximally entangled states like
the Bell states (\ref{be}) because the entangled degree of a quantum
state is invariant under local unitary transformations  in quantum
information \cite{nielsen}.

\section{Quantum algebra associated with the $B_+(x)$-matrix }

The $B_\pm(x)$-matrix (\ref{yb}) is obtained via Yang--Baxterization
of the Bell matrix $B_\pm$, see \cite{yong1,yong2} for the detail,
and it is a solution of the Yang--Baxter equation with the spectral
parameter $x$. There also exists a quantum algebra from the
following $\check{R}TT$ relation: \eq \label{xrtt}
\check{R}(xy^{-1}) (T(x)\otimes T(y))= (T(y)\otimes T(x))\check{R}(x
y^{-1}) \en which is invariant under the rescaling transformation of
the $\check{R}(x)$-matrix and $T(x)$-matrix by global scalar
factors. Here the $\check{R}(x)$-matrix has the form of the
$B_\pm(x)$-matrix (\ref{yb}) without the normalization factor.
Assume the
 $T_\pm(x)$-matrix to have a formalism similar to $B_\pm(x)$ which
 is a linear combination between the Bell matrix $B_\pm$ (\ref{bell})
 and its inverse $B_\pm^{-1}$, \eq B_\pm (x)= B_\pm + 2 x B_\pm^{-1},\qquad
T_\pm(x)= T_\pm + 2 x T_\pm^\prime, \en where the $B_\pm$ matrix
does not have the normalization factor $1/ {\sqrt{2}}$ in
(\ref{bell}), the $T_\pm$-matrix has four noncommutative operators
$\ha,\hb,\hc,\hd$ as its matrix entries and the
$T_\pm^\prime$-matrix has four noncommutative operators
$\ha^\prime,\hb^\prime,\hc^\prime,\hd^\prime$ as its matrix entries.
The original $\check{R}TT$ relation (\ref{xrtt}) with the spectral
parameters $x,y$ is simplified into the four $\check{R}TT$ relations
independent of the spectral parameter,
 \eqa
& & B_\pm (T_\pm\otimes T_\pm) = (T_\pm\otimes T_\pm) B_\pm, \qquad
B_\pm (T_\pm^\prime \otimes T_\pm^\prime) = (T_\pm^\prime \otimes
T_\pm^\prime) B_\pm, \nonumber\\ & &  B_\pm (T_\pm \otimes
T_\pm^\prime) = (T_\pm^\prime \otimes T_\pm) B_\pm,\qquad  B_\pm
(T_\pm^\prime\otimes T_\pm) = (T_\pm\otimes T_\pm^\prime) B_\pm.
 \ena
In the following, for simplicity, we only consider the quantum
algebra determined by the $\check{R}TT$ relations in terms of the
$B_+$ matrix, $T_+$ matrix and $T_+^\prime$ matrix.

Obviously the generators $\ha,\hb,\hc,\hd$ and
$\ha^\prime,\hb^\prime, \hc^\prime,\hd^\prime$ satisfy the same
quantum algebra ${\cal A}_{-1}$. The algebraic equation
 $ B_+(T_+\otimes T_+^\prime)= (T_+^\prime \otimes T_+ ) B_+$ leads to the following
 algebraic relations: \eqa \label{b1}
 & &[a, a^\prime]  = -q c c^\prime- q^{-1} b^{\prime} b,
  \qquad \{a, b^\prime\}=a^\prime b -q c d^\prime, \nonumber\\
  & &\{a, c^\prime\} = c a^\prime + q^{-1} d^\prime b,
  \qquad\,\,\,\,\, [a, d^\prime] = c b^\prime- c^\prime b,  \nonumber\\
  & &[b, a^\prime ] = b^\prime a -q d c^\prime,
 \qquad\qquad\,\, [b,b^\prime ]=-q d d^\prime+ q a^\prime a, \nonumber\\
 & &\{b, c^\prime\} = d a^\prime-d^\prime a,
 \qquad\qquad  \{b, d^\prime\}=d b^\prime-q c^\prime a, \nonumber\\
  & &[c,a^\prime] = -a c^\prime-q^{-1} b^\prime d,
  \qquad\,\, \{c, b^\prime\}=a^\prime d- a d^\prime, \nonumber\\
  & &[c, c^\prime ] = q^{-1} a a^\prime - q^{-1} d^\prime d,
  \qquad \{ c, d^\prime\} = c^\prime d + q^{-1} a b^\prime, \nonumber\\
  & &[d, a^\prime ] = b^\prime c-b c^\prime,
  \qquad\qquad\,\,\,\,\,\,\,\, [d, b^\prime ]= q a^\prime c - b d^\prime,  \nonumber\\
  & &[d, c^\prime ] = d^\prime c + q^{-1} b a^\prime,
  \qquad\qquad [d, d^\prime ]=q^{-1} b b^\prime + q c^\prime c,
 \ena
while the algebraic equation $B_+ (T_+^\prime \otimes T_+)= (T_+
\otimes T_+^\prime) B_+$ leads to more constraint algebraic
relations:
 \eqa \label{b2}
 & & [a, a^\prime ] =q^{-1} b b^\prime + q c^\prime c,
 \qquad[a,b^\prime ]=-b a^\prime + q d^\prime c, \nonumber\\
 & & [a,c^\prime ]= q^{-1} b d^\prime + a^\prime c,
  \,\,\qquad[a, d^\prime ]=-b c^\prime +b^\prime c, \nonumber\\
 & & \{b, a^\prime\} = a b^\prime - q c^\prime d,
\,\,\,\,\,\qquad [b,b^\prime]=-q a a^\prime +q d^\prime d, \nonumber\\
 & &\{b, c^\prime\} = a d^\prime-a^\prime d,
 \,\,\,\,\,\,\,\,\qquad[b,d^\prime]=-q a c^\prime + b^\prime d, \nonumber\\
 & &\{c, a^\prime\} = q^{-1} d b^\prime + c^\prime a,
 \qquad\{c, b^\prime\}=-d a^\prime+ d^\prime a, \nonumber \\
 & &[ c,c^\prime ] = q^{-1} d d^\prime-q^{-1} a^\prime a,
 \,\,\,\,\,\,[c,d^\prime]=-d c^\prime-q^{-1} b^\prime a,\nonumber\\
 & &[d, a^\prime] = c b^\prime-c^\prime b,
\qquad \qquad \{d, b^\prime\} =-q c a^\prime +d^\prime b, \nonumber\\
 & &\{d, c^\prime\} = c d^\prime + q^{-1} a^\prime b,
\qquad [d,d^\prime]=-q c c^\prime-q^{-1} b^\prime b.
 \ena

The algebraic relations (\ref{b1}) determined by $B_+ ( T_+\otimes
T_+^\prime) = (T_+^\prime \otimes T_+ ) B_+$ have the simplified
forms:
 \eqa
 & & [\ha, \ha^\prime] = -q \hc \hc^\prime- q^{-1} \hb^{\prime}\hb,
  \qquad [\hb,\hb^\prime ]=-q \hd \hd^\prime+ q \ha^\prime \ha, \nonumber\\
 & & [\ha, \ha^\prime] =[\hd^\prime, \hd],
   \qquad\qquad\qquad [\hb,\hb^\prime]=q^2 [\hc, \hc^\prime ],\nonumber\\
 & & \{a, b^\prime\}=a^\prime b -q c d^\prime,
   \qquad\qquad [b, a^\prime ]= b^\prime a -q d c^\prime, \nonumber\\
 & & \{a, b^\prime\} =q  [d, c^\prime],
  \qquad\qquad\,\,\,\,\,\,\,[b,a^\prime]=-q \{d^\prime, c\}   \nonumber\\
 & &\{\ha, \hc^\prime\} = \hc \ha^\prime + q^{-1} \hd^\prime \hb,
   \qquad\,\,\,\,  [\hc,\ha^\prime] = -\ha \hc^\prime-q^{-1} \hb^\prime \hd,\nonumber\\
 & &   \{\ha, \hc^\prime\}=q^{-1} [\hd, \hb^\prime ],
   \qquad \qquad \, [\hc,\ha^\prime]=-q^{-1} \{\hb, \hd^\prime\} \nonumber\\
 & & [\ha, \hd^\prime]= \hc \hb^\prime- \hc^\prime \hb,
   \qquad\qquad\,\, \{\hb, \hc^\prime\} = \hd \ha^\prime-\hd^\prime \ha, \nonumber\\
 & & [\ha, \hd^\prime] =[\ha^\prime,\hd],
   \qquad\qquad\,\,\,\,\,\,\,\, \{\hb, \hc^\prime\}= \{\hc, \hb^\prime\}
 \ena
and those algebraic relations (\ref{b2}) from $\check{R} (T_+^\prime
\otimes T_+)= (T_+ \otimes T_+^\prime) \check{R}$ can be also
simplified, \eqa
  & & [\ha, \ha^\prime ] =q^{-1} \hb \hb^\prime + q \hc^\prime \hc,
  \qquad [\hb,\hb^\prime]=-q \ha \ha^\prime +q \hd^\prime \hd, \nonumber\\
 & &  [\ha, \ha^\prime] =[\hd^\prime, \hd],
   \qquad\,\,\,\,\,\,\,\qquad  [\hb,\hb^\prime]=q^2 [\hc,\hc^\prime], \nonumber\\
 & & [\ha,\hb^\prime ]=-\hb \ha^\prime + q \hd^\prime \hc,
  \qquad\,\, \{\ha^\prime, \hb \} = \ha \hb^\prime - q \hc^\prime \hd,\nonumber\\
 & & [\ha, \hb^\prime] = q \{\hd, \hc^\prime\},
 \qquad \qquad \{\ha^\prime, \hb \}=q[\hd^\prime, \hc],  \nonumber\\
 &  &  [\ha,\hc^\prime ]= q^{-1} \hb \hd^\prime + \ha^\prime \hc,
  \qquad \{\hc, \ha^\prime\} = q^{-1} \hd \hb^\prime + \hc^\prime \ha,\nonumber\\
 & &  [\ha,\hc^\prime ]= q^{-1} \{\hd, \hb^\prime\},
  \qquad\,\,\,\,\,  \{\hc, \ha^\prime\}=-q^{-1} [\hb,\hd^\prime],  \nonumber\\
 & & [\ha, \hd^\prime ]=-\hb \hc^\prime +\hb^\prime \hc,
 \qquad\,\,\,\,\,\, \{\hb, \hc^\prime\} = \ha \hd^\prime-\ha^\prime \hd, \nonumber\\
  & & [\ha, \hd^\prime]=[\ha^\prime, \hd],
 \qquad\qquad\,\,\,\,\, \{\hb, \hc^\prime\}=\{\hc, \hb^\prime\}. \ena
 After some further algebraic reductions, two types of generators $\ha, \hb, \hc,\hd$
 and $\ha^\prime, \hb^\prime, \hc^\prime,\hd^\prime$ of the quantum algebra ${\cal B}_{-1}$
 are found to satisfy commutative relations,
\eqa
 & &  \ha\ha^\prime=\ha^\prime\ha,\qquad
 \hd\hd^\prime=\hd^\prime\hd,\qquad \hb \hb^\prime =\hb^\prime\hb,\qquad
 \hc\hc^\prime =\hc \hc^\prime,
 \nonumber\\
  && \ha\hb^\prime =  \ha^\prime\hb,\qquad \hb^\prime\ha=
 \hb\ha^\prime,\qquad \hd^\prime \hc=\hd \hc^\prime,\qquad
 \hc\hd^\prime= \hc^\prime \hd \nonumber\\
 && \ha \hc^\prime = \ha^\prime\hc,\qquad  \hc^\prime\ha =\hc \ha^\prime,\qquad
 \hb \hd^\prime =\hb^\prime \hd,\qquad \hd \hb^\prime =\hd^\prime\hb,
 \nonumber\\
  && \ha^\prime\hd =\ha\hd^\prime,\qquad  \hd\ha^\prime=\hd^\prime\ha, \qquad
  \hb\hc^\prime =\hb^\prime\hc,\qquad  \hc\hb^\prime=\hc^\prime\hb,
 \ena
and additional algebraic relations similar to those (\ref{commu})
determining  ${\cal A}_{-1}$, \eq
   \ha^\prime\ha=\hd\hd^\prime,\,\, \ha \hd^\prime = \hd^\prime
   \ha,\,\, \hb\hb^\prime =- q^2 \hc^\prime \hc,\,\,
 \hb \hc^\prime =-\hc^\prime \hb,\,\,  \ha^\prime \hb = q \hd^\prime
 \hc,\,\, \hb \ha^\prime =-q \hc \hd^\prime, \en
where the quantum algebra determined by theses relations is called
the algebra ${\mathcal B}_{-1}$ and the deformation parameter $q$ is
irrelevant due to the rescaling transformation: $q\hc\to\hc$ and
$q\hc^\prime\to\hc^\prime$. Explicitly, further research will be
needed to discover interesting algebraic structures underlying the
quantum algebra ${\cal B}_{-1}$.

\section{Concluding remarks and problems}

In this paper, we derive the quantum algebra ${\cal A}_{-1}$ using
the $FRT$ construction of the Bell matrix, and list its two
dimensional representations and pick up two characteristic examples
for its four dimensional representations. The first example has a
composition series over the complex field $\mathbb C$ and the second
leads to $\underline{2}\otimes \underline{2}=\underline{2}\oplus
\underline{2}$ which is not true for the representation of the Lie
algebra \cite{kassel}. In addition, we have the quantum algebra
determined by the $\check{R}TT$ relation of Yang--Baxterization of
the Bell matrix.

Besides these topics in the present paper and \cite{wang, yong5},
there still remain a series of problems to be solved in the project
of exploring algebraic structures associated with the Bell matrix.
In view of known achievements in quantum groups \cite{kassel},
readers are invited to study the following problems:\\[-.8cm]

 \begin{enumerate}

  \item Does ${\cal A}_{-1}$\footnote{Note that the quantum algebra ${\cal
  A}_{-1}$ is not the deformation of  classical algebras of functions on the
  Lie algebra because the Bell-matrix is not a deformation of the trivial
  $\check{R}$-matrix which is the identity up to signs.}
  have a center like the quantum determinant \cite{kassel} and a four-dimension
  representation: $\underline{2}\otimes \underline{2}=\underline{1}\oplus
  \underline{3}$?    \\[-.7cm]

 \item The construction of universal $\check{R}$-matrix \cite{kassel}
 in terms of generators of ${\cal A}_{-1}$. \\[-.7cm]

  \item Interesting algebraic structures underlying ${\cal B}_{-1}$ such
  as the quantum double.\\[-.7cm]

 \item  New quantum algebras obtained by exploiting methodologies for
 the Sklyanin algebra \cite{sklyanin1,sklyanin2} to the Bell matrix.\\[-.7cm]

 \item Physical realization of ${\cal A}_{-1}$ (or ${\cal B}_{-1}$)
  and its application.\\[-.7cm]

\end{enumerate}

{\em Note added.} After this paper is submitted into the web, the
authors are informed that the quantum algebra (\ref{algebra}) has
been already presented by Arnaudon, Chakrabarti, Dobrev and Mihov,
see \cite{acdm1,acdm2,acdm3} which study the exotic bialgebras
obtained by the $FRT$ recipe on the non-triangular non-singular
$R$-matrix in view of the classification of the constant
Yang--Baxter solution by Hietarinta \cite{hietarinta}. As the
authors of \cite{acdm1,acdm2,acdm3} have admitted during email
correspondences between two groups, our research is completely
independent of \cite{acdm1,acdm2,acdm3}. Here it is still necessary
to make essential differences clear between both research projects.
(1) {\it Motivations.} The articles \cite{acdm1,acdm2,acdm3} aim to
finalize the explicit classification of the matrix bialgebras
generated by four elements. This paper is completely motivated by
the recent study of quantum information
\cite{dye,kauffman,yong1,yong2,yong3,yong4}, and it sheds a light on
the connection between quantum information and quantum groups,
therefore it suggests a new interdisciplinary field which can be
appreciated by the community of quantum groups. (2) {\it Quantum
algebras.} The articles \cite{acdm1,acdm3} use the $RTT$ relation to
derive the quantum algebra called $S03$ and our paper exploits
$\check{R}TT$ relation to obtain the quantum algebra ${\cal
A}_{-1}$. The $\check{R}$-matrix in the present paper is called the
Bell matrix which is a unitary braid representation, but the
$R$-matrix \cite{acdm1,acdm2,acdm3} can not be called the Bell
matrix since it is not a solution of the braided Yang--Baxter
equation. Also, the $\check{R}$-matrix has the deformation parameter
$q$ and  ${\cal A}_{-1}$ is proved to be independent of $q$.
Furthermore, ${\mathcal A}_{-1}$ is determined by six algebraic
relations but $S03$ \cite{acdm1,acdm3}, derived from the $RTT$
relation, has eight algebraic equations. In this sense, we actually
have a ``new" quantum algebra ${\cal A}_{-1}$ which has the quotient
algebra $S03$. (3) {\it Representations.} It is not very easy to
compare representations of quantum algebras given by
\cite{acdm1,acdm2,acdm3}  and the present paper due to the fact that
the style, language and notation taken by both groups are very
different. The article  \cite{acdm1} shows representations of the
dual bialgebra $s03$ of $S03$ instead of representations of $S03$.
Our paper presents almost all two dimensional representations of
${\cal A}_{-1}$. More essentially, the article \cite{acdm1} focuses
on the representation theory of the dual bialgebra $s03$ but ours is
looking for interesting algebraic structures in the representation
theory of ${\cal A}_{-1}$. For example, we have obtained composition
series,
$\underline{2}\otimes\underline{2}=\underline{2}\oplus\underline{2}$,
cyclic representations, and a representation formed by four
maximally entangled states which are local unitary transformations
of Bell states.  (4) {\it $FRT$ dual algebra.} The article
\cite{acdm2} derives the $FRT$ bialgebra $s03_F$ of $S03$ and gives
its representations instead of representations of $S03$. The present
paper also obtains the $FRT$ dual algebra without studying its
representations. But these algebraic relations are not completely
the same which both groups exploit to derive the $FRT$ dual algebra.
Besides calculation detail, we show the algebraic relations defining
the $FRT$ dual algebra in a clear way, find the mixed algebraic
relations similar to those determining ${\cal A}_{-1}$ and denote
this algebra by ${\cal B}_{-1}$. (5) {\it Physical applications.}
The article \cite{acdm3}  considers an exotic eight-vertex model and
an integrable spin-chain model. But the present paper suggests in
its introduction that the Bell matrix and its Yang--Baxterization
have negative entries which can not be explained as positive (zero)
Boltzman weights in statistical physics, and therefore the authors
prefer physical applications of the Bell matrix, quantum algebras
${\cal A}_{-1}$ and ${\cal B}_{-1}$ to quantum information theory.
For example, see \cite{kauffman,yong1,yong2,yong3,yong4}, in quantum
computation the Bell matrix $B_+$ ($B_{-}$) can be identified as a
universal quantum gate and the permutation $P$ matrix is the swap
gate, hence the $R$-matrix \cite{acdm1,acdm2,acdm3} can also
recognized as a universal quantum gate $R=B_+P$ ($R=PB_{-}$).
Moreover, $P$, $B_{+}$ ($B_-$) can generate the virtual braid group
which is a natural language for topological quantum computing
\cite{yong5,yong6}, i.e., the $R$-matrix \cite{acdm1,acdm2,acdm3} is
an element of the virtual braid group. (6) {\it Geometry.} The
article \cite{acdm2}  briefly discusses the associated
noncommutative geometry with their $R$-matrix but this paper does
not study it.

 \section*{Acknowledgments}

  The authors thank A. Chakrabarti and V.K. Dobrev for their helpful
  references and comments on this manuscript. Y. Zhang thanks K. Fujii,
  L.H. Kauffman and M.A. Martin-Delgado for their helpful suggestions
  on further research, and he thanks partial supports from NSFC grants
  and SRF for ROCS, SEM. N. Jing thanks Chern Institute of Mathematics
  for the hospitality during his visiting period and thanks partial support
  from NSA grant.

\end{document}